\documentstyle[multicol,aps,psfig]{revtex}
\begin{document}
\preprint{LBNL-49561}

\title{Jet Tomography of Hot and Cold Nuclear Matter}
\author{Enke Wang$^{a,b}$ and Xin-Nian Wang$^{b,a}$}
\address{$^a$ Institute of Particle Physics, Huazhong Normal University,
         Wuhan 430079, China}
\address{$^b$Nuclear Science Division, MS 70-319,
Lawrence Berkeley National Laboratory, Berkeley, CA 94720 USA}

\date{January 28, 2002}

\maketitle

\vspace{-1.5in}
{\hfill LBNL-49561}
\vspace{1.4in}

\begin{abstract}
Medium modification of parton fragmentation functions induced
by multiple scattering and gluon bremsstrahlung 
is shown to describe the recent HERMES data in
deeply inelastic scattering (DIS) very well, providing the first
evidence of $A^{2/3}$-dependence of the modification. 
The energy loss is found to be $\langle dE/dL\rangle \approx 0.5$ GeV/fm 
for a 10-GeV quark in a $Au$ nucleus.
Including the effect of expansion, analysis of the $\pi^0$ spectra 
in central $Au+Au$ collisions at $\sqrt{s}=130$ GeV yields an 
averaged energy loss equivalent to $\langle dE/dL\rangle \approx 7.3$ GeV/fm 
in a static medium.
Predictions for central $Au+Au$ collisions 
at $\sqrt{s}=200$ GeV are also given.

\end{abstract}

\begin{multicols}{2}

Energetic partons produced via hard processes 
provide an excellent tool that enables tomographic studies of both hot 
dense and cold nuclear matter. By measuring the 
attenuation \cite{gw1,bdms,zhak,glv,wied} of these partons as they 
propagate through the medium, one would be able to study the 
properties such as the geometry \cite{wv2,gvw} and the gluon density 
of the medium. The attenuation will suppress the final 
leading hadron distribution giving rise to
modified parton fragmentation functions \cite{wh}.
Such a modified fragmentation function inside a nucleus has been derived 
recently \cite{gw01} in a pQCD approach with a systematic
expansion of higher-twist corrections to the fragmentation processes.
In this letter, we will compare the predicted nuclear
modification to the recent HERMES experimental data \cite{hermes} and 
extract the effective parton energy loss. We will 
further extend the study to the case of a hot QCD medium
including the dynamics of expansion. We will then analyze within this 
framework the $\pi^0$ spectra as measured by the PHENIX 
experiment \cite{phenix} in central $Au+Au$ collisions at $\sqrt{s}=130$ GeV,
which have shown significant suppression at large transverse momentum. The
extracted effective parton energy loss will be compared with that in a cold
nucleus, and discussions will be given about the implications of the
PHENIX data on the gluon density in the early stage of central
$Au+Au$ collisions at the RHIC energy. We will also provide predictions
for $\pi^0$ spectra in central $Au+Au$ collisions at $\sqrt{s}=200$ GeV.

In deeply inelastic scatterings (DIS),
 quark fragmentation functions are factorizable in leading twist
from the parton distribution functions and photon-quark
($\gamma^*q$) scattering cross section. In a nucleus target, the quark 
struck by the virtual photon suffers multiple scattering
and induced bremsstrahlung before hadronization. Extending the generalized 
factorization \cite{lqs} to the semi-inclusive process, 
$e(L_1) + A(p) \longrightarrow e(L_2) + h (\ell_h) +X $, one 
can define a modified fragmentation function as \cite{gw01},
\begin{eqnarray}
&&E_{L_2}\frac{d\sigma_{\rm DIS}^h}{d^3L_2dz_h}
=\frac{\alpha^2_{\rm EM}}{2\pi s}\frac{1}{Q^4} L_{\mu\nu} \nonumber \\
&\times&\sum_q \int dx f_q^A(x,Q^2) 
H^{(0)}_{\mu\nu}(x,p,q)
\widetilde{D}_{q\rightarrow h}(z_h,Q^2)\; , \label{eq:Wtot}
\end{eqnarray}
where $f_q^A(x,Q^2)$ is the quark distribution function in the nucleus,
$s=(p+L_1)^2$ and $p=[p^+,0,\vec{0}_\perp]$ is the 
momentum per nucleon inside the nucleus. The momentum of the virtual 
photon $\gamma^*$ is $q=[-Q^2/2q^-,q^-,\vec{0}_\perp]$ and the momentum
fraction carried by the hadron is $z_h=\ell_h^-/q^-$.
The hard part of the $\gamma^*q$ scattering $H_{\mu\nu}^{(0)}$ is the 
same as in $ep$ scattering \cite{gw01} and 
$L_{\mu\nu}=\frac{1}{2}\, {\rm Tr}(\not{\hspace{-3pt}L_1} \gamma_{\mu}
\not{\hspace{-3pt}L_2} \gamma_{\nu})$.

Including the leading twist-4 contributions from double scattering processes,
the modified effective quark fragmentation function can be 
obtained as \cite{gw01}
\begin{equation}
\widetilde{D}_{q\rightarrow h}(z_h,Q^2) \equiv
D_{q\rightarrow h}(z_h,Q^2) + \Delta D_{q\rightarrow h}(z_h,Q^2);
\end{equation}
\end{multicols}
\begin{equation}
\Delta D_{q\rightarrow h}(z_h,Q^2)= 
\int_0^{Q^2} \frac{d\ell_T^2}{\ell_T^2} 
\frac{\alpha_s}{2\pi} \int_{z_h}^1 \frac{dz}{z}
\left[ \Delta\gamma_{q\rightarrow qg}(z,x,x_L,\ell_T^2) 
D_{q\rightarrow h}(z_h/z)
+ \Delta\gamma_{q\rightarrow gq}(z,x,x_L,\ell_T^2)
D_{g\rightarrow h}(z_h/z)\right] \, , \label{eq:dmod}
\end{equation}
\begin{equation}
\Delta\gamma_{q\rightarrow qg}(z,x,x_L,\ell_T^2)=
\left[\frac{1+z^2}{(1-z)_+}T^A_{qg}(x,x_L) + 
\delta(1-z)\Delta T^A_{qg}(x,\ell_T^2) \right]
\frac{C_A2\pi\alpha_s}
{(\ell_T^2+\langle k_T^2\rangle)N_c f_q^A(x,\mu_I^2)} \; ,
\end{equation}
where, $C_A=3$, $N_c=3$,
$\Delta\gamma_{q\rightarrow gq}(z,x,x_L,\ell_T^2) 
= \Delta\gamma_{q\rightarrow qg}(1-z,x,x_L,\ell_T^2)$
are the modified splitting functions,
$x_L=\ell_T^2/2p^+q^-z(1-z)$, and $D_{a\rightarrow h}(z_h,Q^2)$ 
are the normal twist-2 parton fragmentation functions in vacuum.
The $\delta$-function part in the modified splitting function is 
from the virtual corrections, with $\Delta T^A_{qg}(x,\ell_T^2)$ 
defined as
\begin{equation}
\Delta T^A_{qg}(x,\ell_T^2) \equiv
\int_0^1 dz\frac{1}{1-z}\left[ 2 T^A_{qg}(x,x_L)|_{z=1}
-(1+z^2) T^A_{qg}(x,x_L)\right] \, . \label{eq:vsplit}
\end{equation}
Such virtual or absorptive corrections are important to 
ensure the infrared safety of the modified fragmentation function 
and the unitarity of the gluon radiation processes. 
The quark-gluon correlation function 
\begin{eqnarray}
T^A_{qg}(x,x_L)&=& \int \frac{dy^{-}}{2\pi}\, dy_1^-dy_2^-
e^{i(x+x_L)p^+y^-+ix_Tp^+(y^-_1-y^-_2)}(1-e^{-ix_Lp^+y_2^-})
(1-e^{-ix_Lp^+(y^--y_1^-)}) \nonumber \\
&\frac{1}{2}&\langle A | \bar{\psi}_q(0)\,
\gamma^+\, F_{\sigma}^{\ +}(y_{2}^{-})\, F^{+\sigma}(y_1^{-})\,\psi_q(y^{-})
| A\rangle \theta(-y_2^-)\theta(y^--y_1^-)
\label{eq:qgmatrix}
\end{eqnarray}
\begin{multicols}{2}
\noindent contains essentially four independent twist-4 parton matrix elements 
in a nucleus [$x_T=\langle k_T^2\rangle/2p^+q^-z(1-z)$].
The dipole-like form-factor $(1-e^{-ix_Lp^+y_2^-})(1-e^{-ix_Lp^+(y^--y_1^-)})$
arises from the interference between the final state radiation of 
the $\gamma^*q$ scattering and the gluon bremsstrahlung induced
by the secondary quark-gluon scattering. By generalizing the 
factorization assumption \cite{lqs} to these twist-four
parton matrices, we have
\begin{equation}
T^A_{qg}(x,x_L)\approx \widetilde{C}(Q^2)
m_NR_A f_q^A(x) (1-e^{-x_L^2/x_A^2}),
\label{eq:tqg2}
\end{equation}
with a Gaussian nuclear distribution 
$\rho(r)\sim \exp(-r^2/2R_A^2)$, $R_A=1.12 A^{1/3}$ fm.
Here, $x_A=1/m_NR_A$, and $m_N$ is the nucleon mass.
We should emphasize that the parameter  $\widetilde{C}(Q^2)$ should 
in principle depend on the renormalization scale $Q^2$ among
other kinetic variables as shown recently in a detailed analysis of
the twist-four nuclear matrix elements \cite{ow}. Therefore, it can take
different values in different processes.

\begin{figure}
\centerline{\psfig{figure=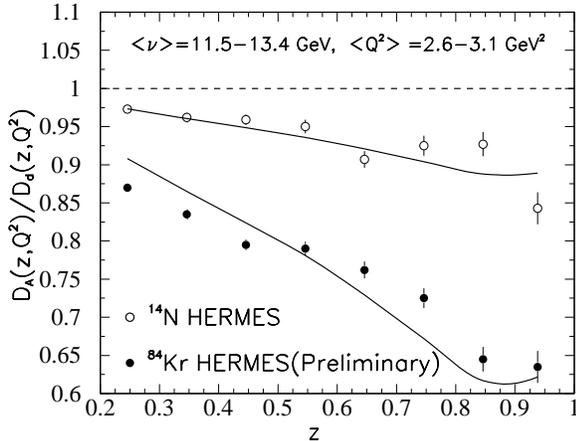,width=3.0in,height=2.3in}}
\caption{Predicted nuclear modification of jet fragmentation function
is compared to the HERMES data \protect\cite{hermes} on ratios of
hadron distributions between $A$ and $D$ targets in DIS.}
\label{fig1}
\end{figure}

Since the two interference terms in the dipole-like form-factor
involve transferring momentum $x_Lp^+$ between different nucleons 
inside a nucleus, they should be suppressed for large nuclear size 
or large momentum fraction $x_L$. Notice that $\tau_f=1/x_Lp^+$ is 
the gluon's formation time. Thus, $x_L/x_A=L_A/\tau_f$, with $L_A=R_Am_N/p^+$ 
being the nuclear size in the infinite momentum frame.
The effective parton correlation and the induced gluon emission 
vanishes when the formation time is much larger than the nuclear size,
$x_L/x_A\ll 1$, because of the Landau-Pomeranchuck-Migdal (LPM) 
interference effect.
Therefore, the LPM interference restricts the radiated gluon to have 
a minimum transverse momentum $\ell_T^2\sim Q^2/m_NR_A\sim Q^2/A^{1/3}$. 
The nuclear corrections to the fragmentation function due to double 
parton scattering will then be in the order of 
$\alpha_s A^{1/3}/\ell_T^2 \sim \alpha_s A^{2/3}/Q^2$, which depends
quadratically on the nuclear size. For large values of
$A$ and $Q^2$, these corrections are leading; yet the requirement
$\ell_T^2\ll Q^2$ for the logarithmic approximation in deriving the 
modified fragmentation function is still valid.

\begin{figure}
\centerline{\psfig{figure=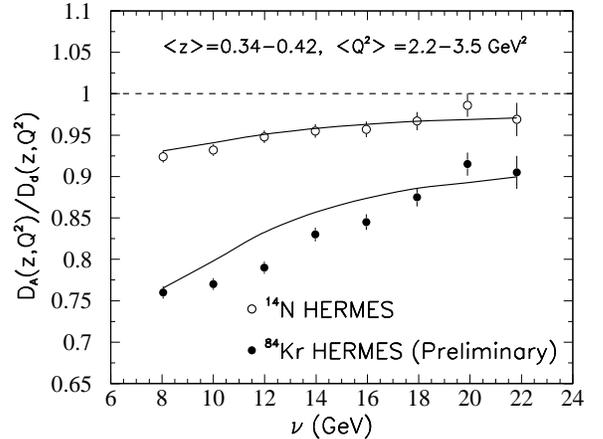,width=3.0in,height=2.3in}}
\caption{Energy dependence of the nuclear modification compared
with the HERMES data \protect\cite{hermes}.}
\label{fig2}
\end{figure}

With the assumption of the factorized form of the twist-4
nuclear parton matrices, there is only one free parameter 
$\widetilde{C}(Q^2)$
which represents quark-gluon correlation strength inside nuclei.
Once it is fixed, one can predict the $z$, energy and
nuclear dependence of the medium modification of the fragmentation
function. Shown in Figs.~\ref{fig1} and \ref{fig2} are the calculated 
nuclear modification factor of the fragmentation functions for $^{14}N$ 
and $^{84}Kr$ targets as compared to the recent HERMES data \cite{hermes}.
There are strong correlations among values of $Q^2$, $\nu$ and $z$ in the
HERMES data which are also taken in account in our calculation.
The predicted shape of the $z$- and $\nu$-dependence agrees well 
with the experimental data.  A remarkable feature of the prediction
is the quadratic $A^{2/3}$ nuclear size dependence, which is verified 
for the first time by an experiment. This quadratic dependence comes 
from the combination of the QCD radiation spectrum and the modification 
of the available phase space in $\ell_T$ or $x_L$ due to the LPM 
interferences.

The observed attenuation of the leading hadrons could also be attributed 
phenomenologically to hadron absorption inside the nucleus. This is, 
however, achieved only via some {\it ad hoc} assumption of the hadron 
formation time \cite{hermes}. Considering the hadronization process as 
regeneration of gluon field within a spatial region of a hadon 
size $r_h$, the hadron formation time for a light quark
 will be $t_f^h \approx \nu r_h^2$ \cite{dkmt}.
Taking $r_h \approx 1$ fm, $t_f^h$ is about 40 fm 
for $\nu=8$ GeV, the lower limit of the HEREMS experiment. 
This is much larger than the size of the heaviest nuclei available. 
We therefore assume that the nuclear attenuation in this energy region 
is mainly caused by induced gluon radiation and multiple parton scattering.

By fitting the overall suppression for one nuclear target, 
we obtain the only parameter in our calculation,
$\widetilde{C}(Q^2)=0.0060$ GeV$^2$ 
with $\alpha_{\rm s}(Q^2)=0.33$ at $Q^2\approx 3$ GeV$^2$.
This parameter is also related to nuclear broadening of 
the transverse momentum of the Drell-Yan dilepton in 
$pA$ collisions \cite{guo},
$\langle\Delta q_\perp^2\rangle\approx\widetilde{C}\pi\alpha_{\rm s}/N_cx_A$.
With an experimental value of 
$\langle\Delta q_\perp^2\rangle=0.016A^{1/3}$ GeV$^2$ \cite{peng} and
$\alpha_{\rm s}(M^2_{l\bar{l}})=0.21$ ($<M_{l\bar{l}}^2>\approx 40$ GeV$^2$),
one finds $\widetilde{C}(M^2_{l\bar{l}})=0.013$ GeV$^2$, 
which is about a factor 2
larger than the value obtained in our fit to the HERMES data. The value of 
$\widetilde{C}$ determined from nuclear broadening in photo-production
of di-jet is even larger \cite{lqs} at $Q^2=4p_T^2\approx 64$ GeV$^2$.
Such a strong scale dependence of $\widetilde{C}(Q^2)$ 
is in line with one's expectation
since it is related to gluon distribution $xg(x,Q^2)$ at 
small $x$ in nuclei \cite{ow}.

If one defines theoretically the quark energy loss as that carried
by the radiated gluons, then the averaged total fractional
energy loss is [cf. Eq.~(\ref{eq:dmod})],
\begin{eqnarray}
&&\langle\Delta z_g\rangle 
=\int_0^{\mu^2}\frac{d\ell_T^2}{\ell_T^2} 
\int_0^1 dz \frac{\alpha_s}{2\pi}
 z\,\Delta\gamma_{q\rightarrow gq}(z,x_B,x_L,\ell_T^2) \nonumber \\
&=& \int_0^{Q^2}d\ell_T^2 \int_0^1 dz
\frac{1+(1-z)^2}{\ell_T^2(\ell_T^2+\langle k_T^2\rangle)} 
\frac{C_A\alpha_s^2T_{qg}^A(x_B,x_L)}{N_cf_q^A(x_B)} \\
&\approx&\widetilde{C}(Q^2)\alpha_{\rm s}^2(Q^2) \frac{C_A}{N_c} \frac{x_B}{Q^2x_A^2}
6\ln\left(\frac{1}{2x_B}\right)\; .
 \label{eq:loss1}
\end{eqnarray}
In the rest frame of the nucleus, $p^+=m_N$, $q^-=\nu$, and
$x_B\equiv Q^2/2p^+q^-=Q^2/2m_N\nu$. One can
get the averaged total energy loss as
$ \Delta E=\nu\langle\Delta z_g\rangle
\approx  \widetilde{C}(Q^2)\alpha_{\rm s}^2(Q^2)
m_NR_A^2(C_A/N_c) 3\ln(1/2x_B)$.
With the determined value of $\widetilde{C}$, 
$\langle x_B\rangle \approx 0.124$ in the HERMES experiment \cite{hermes}
and the average distance $\langle L_A\rangle=R_A\sqrt{2/\pi}$
for the 
assumed Gaussian nuclear distribution,
one gets the quark energy 
loss $dE/dL\approx 0.5$ GeV/fm inside a $Au$ nucleus.

To extend our study of modified fragmentation functions to 
jets in heavy-ion collisions, we 
assume $\langle k_T^2\rangle\approx \mu^2$ (the Debye screening mass)
and a gluon density profile
$\rho(y)=(\tau_0/\tau)\theta(R_A-y)\rho_0$ for a 1-dimensional 
expanding system. Since the initial 
jet production 
rate is independent of the final gluon
density 
which can be related to the 
parton-gluon scattering cross section \cite{bdms} [
$\alpha_s x_TG(x_T)\sim \mu^2\sigma_g$], one has then
\begin{equation}
\frac{\alpha_s T_{qg}^A(x_B,x_L)}{f_q^A(x_B)} \sim
\mu^2\int dy \sigma _g \rho(y)
[1-\cos(y/\tau_f)],
\end{equation}
where $\tau_f=2Ez(1-z)/\ell_T^2$ is the gluon formation time. One
can recover the form of energy loss in a thin plasma obtained 
in the opacity expansion approach \cite{bdms,gvw},
\begin{eqnarray}
\langle\Delta z_g\rangle &=&\frac{C_A\alpha_s}{\pi}
\int_0^1 dz \int_0^{\frac{Q^2}{\mu^2}}du \frac{1+(1-z)^2}{u(1+u)} \nonumber \\
&\times&\int_{\tau_0}^{R_A} d\tau\sigma_g\rho(\tau) 
\left[1-\cos\left(\frac{(\tau-\tau_0)\,u\,\mu^2}{2Ez(1-z)}\right)\right].
\end{eqnarray}
Keeping only the dominant contribution and assuming 
$\sigma_g\approx C_a 2\pi\alpha_s^2/\mu^2$ ($C_a$=1 for $qg$ and 9/4 for
$gg$ scattering), one obtains the averaged energy loss,
\begin{equation}
\langle \frac{dE}{dL}\rangle \approx \frac{\pi C_aC_A\alpha_s^3}{R_A}
\int_{\tau_0}^{R_A} d\tau \rho(\tau) (\tau-\tau_0)\ln\frac{2E}{\tau\mu^2}.
\end{equation}
Neglecting the logarithmic dependence on $\tau$, the averaged energy loss
in a 1-dimensional expanding system can be expressed as
\begin{equation}
\langle\frac{dE}{dL}\rangle_{1d} \approx \frac{dE_0}{dL} \frac{2\tau_0}{R_A},
\end{equation}
where $dE_0/dL\propto \rho_0R_A$
is the energy loss in a static medium with the same gluon density $\rho_0$ 
as in a 1-d expanding system at time $\tau_0$.
Because of the expansion, the averaged energy loss $\langle dE/dL\rangle_{1d}$
is suppressed as compared to the static case and does not depend linearly
on the system size. This could be one of the reasons why the effect of
parton energy loss is found to be negligible in $AA$ collisions at
$\sqrt{s}=17.3$ GeV \cite{wangsps}.

An effective model of modified fragmentation functions was
proposed in Ref.~\cite{wh}:
\begin{equation}
\widetilde{D}_{a\rightarrow h} (z)\approx \frac{1}{1-\Delta z} 
D_{a\rightarrow h}\left(\frac{z}{1-\Delta z}\right), \label{dbar0}
\end{equation}
with $\Delta z$ to account for the fractional parton energy loss. 
This effective model is found to
reproduce the pQCD result 
from Eq.~(\ref{eq:dmod}) very well, but only when
$\Delta z$ is set to be $\Delta z\approx 0.6 \langle z_g\rangle$.
Therefore the actual averaged parton energy loss should be
$\Delta E/E=1.6\Delta z$ with $\Delta z$ extracted from the 
effective model. The factor 1.6 is mainly
caused by unitarity correction effect in 
the pQCD calculation. A similar effect is also found in the 
opacity expansion approach \cite{glv02}.

\begin{figure}
\centerline{\psfig{figure=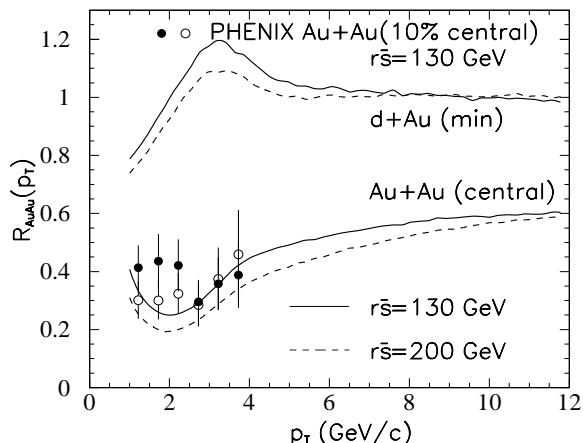,width=3.0in,height=2.3in}}
\caption{Calculated nuclear modification factor of $\pi^0$ $p_T$
spectra for $d+Au$ and central $Au+Au$ collisions at $\sqrt{s}=130$ (solid)
and 200 GeV (dashed) as compared to PHENIX data \protect\cite{phenix}
(solid circles are measured spectra normalized by
PHENIX parameterization of the $pp$ spectra while open circles 
are normalized by our calculated $pp$ spectra).}
\label{fig3}
\end{figure}

The PHENIX experiment has reported \cite{phenix} a strong suppression 
of high $p_T$ hadrons in central $Au+Au$ collisions at $\sqrt{s}=130$ GeV.
To extract the parton energy loss, we compare the data
with the calculated hadron $p_T$ spectra in heavy-ion collisions 
using the above effective model for medium modified jet fragmentation 
functions \cite{wang98}.
Shown in Fig.~\ref{fig3} are the nuclear modification 
factor $R_{AA}(p_T)$ as the ratio of hadron spectra in $AA$ ($pA$) 
and $pp$ collisions normalized by the number of binary collisions\cite{ww}.
Parton shadowing and nuclear broadening of the intrinsic $k_T$ 
are also taken into account in the calculation which decrsibes $pA$
data for energies up to $\sqrt{s}=40$ GeV \cite{wang98}.
The nuclear $k_T$-broadening gives the so-called Cronin 
enhancement at large $p_T$ in $pA$ collisions, where there is no
parton energy loss induced by a hot medium. 
Fitting the PHENIX data 
yields  $\langle dE/dL\rangle_{1d} \approx 0.34\ln E/\ln 5$ GeV/fm, 
including the 
factor of 1.6 from the unitarity correction effect. We consider only
$\pi^0$ data here, since at large $p_T$ the charged hadrons are dominated by
baryons, which could be influenced mainly by non-perturbative
dynamics \cite{gvw}.

Taking into account the expansion, the averaged parton
energy loss extracted from the PHENIX data 
would be equivalent to $(dE/dL)_0=0.34 (R_A/2\tau_0)\ln E/\ln 5$ 
in a static system 
with the same gluon density as the initial value of the
expanding system at $\tau_0$. With $R_A\sim 6$ fm and $\tau_0\sim 0.2$ fm,
this would give $(dE/dL)_0\approx 7.3$ GeV/fm for a 10-GeV parton, 
which is about 15 times of that in a cold $Au$ nucleus as extracted 
from the HERMES data. 
Since the parton energy loss is directly proportional to gluon density of
the medium,
we can predict the $\pi^0$ spectra at $\sqrt{s}=200$ GeV as given by
the dashed lines in Fig.~\ref{fig3}, assuming that the initial parton 
density in central $Au+Au$ collisions at $\sqrt{s}=200$ GeV is 
about 10\% higher than at 130 GeV.

In summary, the nuclear modification of parton fragmentation function
predicted in a pQCD study describes well the HERMES experimental data.
The extracted energy loss is $dE/dL\approx 0.5$ GeV/fm for a quark
with $E=10$ GeV in a $Au$ nucleus. 
Analysis of the PHENIX data of $\pi^0$ spectra 
suppression in central $Au+Au$ collisions yields an averaged parton 
energy loss in an expanding system that would be equivalent to 
$(dE/dL)_0\approx 7.3$ GeV/fm in a static medium. 

This work was supported by DOE under contract No. DE-AC03-76SF00098 
and by NSFC under project No. 19928511 and No. 10135030. We thank
N. Bianchi and P. DiNezza for discussions about the HERMES data.


\end{multicols}


\vspace{-0.1in}
\begin{thebibliography}{99}

\bibitem{gw1}M. Gyulassy and X.-N. Wang, Nucl. Phys. {\bf B}420, 583 (1994);
X.~N.~Wang, M.~Gyulassy and M.~Plumer,
Phys.\ Rev.\ D {\bf 51}, 3436 (1995).

\bibitem{bdms} R. Baier {\it et al.},
Nucl. Phys. {\bf B484}, 265 (1997);
Phys.\ Rev.\ C {\bf 58}, 1706 (1998).

\bibitem{zhak}B. G. Zhakharov, JETP letters {\bf 63}, 952 (1996).

\bibitem{glv} M. Gyulassy, P. L\'evai and I. Vitev, Nucl. Phys. 
         {\bf B594}, 371 (2001); 
Phys. Rev. Lett. {\bf 85}, 5535 (2000).

\bibitem{wied}U. Wiedemann, Nucl. Phys. {\bf B588}, 303 (2000).

\bibitem{wv2} X.-N.~Wang,
Phys.\ Rev.\ C {\bf 63}, 054902 (2001).

\bibitem{gvw}
M.~Gyulassy, I.~Vitev and X.-N.~Wang,
Phys.\ Rev.\ Lett.\  {\bf 86} (2001) 2537;
M.~Gyulassy, I.~Vitev, X.~N.~Wang and P.~Huovinen,
Phys.\ Lett.\ B {\bf 526}, 301 (2002).

\bibitem{wh}
X.-N.~Wang and Z.~Huang,
Phys.\ Rev.\ C {\bf 55} (1997) 3047;
X.-N.~Wang, Z.~Huang and I.~Sarcevic,
Phys.\ Rev.\ Lett.\  {\bf 77} (1996) 231.

\bibitem{gw01}
X.-N.~Wang and X.~F.~Guo,
Nucl.\ Phys.\ A {\bf 696}, 788 (2001);
Phys.\ Rev.\ Lett.\  {\bf 85} (2000) 3591.
\bibitem{hermes}
A.~Airapetian {\it et al.}
, Eur.\ Phys.\ J.\ C {\bf 20}, 479 (2001);
V.~Muccifora 
, arXiv:hep-ex/0106088.

\bibitem{phenix}
K.~Adcox {\it et al.} 
, Phys.\ Rev.\ Lett.\  {\bf 88}, 022301 (2002).

\bibitem{lqs}
 M. Luo, J. Qiu and G. Sterman, Phys. Lett. {\bf B279}, 377 (1992);
Phys. Rev. D{\bf 50}, 1951 (1994).

\bibitem{ow}
J.~Osborne and X.-N.~Wang,
arXiv:hep-ph/0204046.

\bibitem{dkmt}
Yu. L. Dokshitzer, V. A. Khoze, A. H. Mueller and S. I. Troyan,
{\it Basics of Perturbative QCD} (Editions Frontieres, 
Gif-sur-Yvette Cedex, 1991).

\bibitem{guo}
X.~F.~Guo,
Phys.\ Rev.\ D {\bf 58}, 114033 (1998).

\bibitem{peng}
P.~L.~McGaughey, J.~M.~Moss and J.~C.~Peng,
Ann.\ Rev.\ Nucl.\ Part.\ Sci.\  {\bf 49}, 217 (1999);
J.~C. Peng, priviate
communication.



\bibitem{glv02}
M.~Gyulassy, P.~Levai and I.~Vitev,
arXiv:nucl-th/0112071.

\bibitem{wang98}
X.-N.~Wang, Phys.\ Rev.\ C {\bf 61}, 064910 (2000).

\bibitem{ww}
E.~Wang and X.-N.~Wang,
Phys.\ Rev.\ C {\bf 64}, 034901 (2001).

\bibitem{wangsps}
X.-N.~Wang,
Phys.\ Rev.\ Lett.\  {\bf 81}, 2655 (1998).

\end{thebibliography}
\end{document}